\documentclass[twocolumn,prl,floatfix,preprintnumbers,a4paper,nofootinbib,superscriptaddress]{revtex4}

\usepackage{color}
\usepackage{calc}
\usepackage{amsmath,amssymb,graphicx}
\usepackage{physics}
\usepackage{amssymb,amsmath}
\usepackage{tensor}
\usepackage{bm}
\usepackage{microtype}
\usepackage{booktabs}
\usepackage{times}
\usepackage[varg]{txfonts}
\usepackage[colorlinks, pdfborder={0 0 0}]{hyperref}
\usepackage{subfigure}
\definecolor{LinkColor}{rgb}{0.75, 0, 0}
\definecolor{CiteColor}{rgb}{0, 0.5, 0.5}
\definecolor{UrlColor}{rgb}{0, 0, 0.75}
\hypersetup{linkcolor=LinkColor}
\hypersetup{citecolor=CiteColor}
\hypersetup{urlcolor=UrlColor}
\allowdisplaybreaks
\usepackage[utf8]{inputenc}
\usepackage{ulem}
\allowdisplaybreaks

\newcommand{\blambda}{\bm{\lambda}}
\newcommand{\Ms}{M_\mathrm{s}}

\newcommand{\Mo}{M_\mathrm{o}}

\newcommand{\ms}{m_\mathrm{s}}
\newcommand{\mo}{m_\mathrm{o}}
\newcommand{\zs}{z_\mathrm{s}}
\newcommand{\zo}{z_\mathrm{o}}
\newcommand{\Gs}{G_\mathrm{s}}

\newcommand{\Go}{G_\mathrm{o}}

\begin{document}

\title{Constraints on the time variation of the gravitational constant using gravitational-wave observations of binary neutron stars}

\author{Aditya Vijaykumar}
\affiliation{International Centre for Theoretical Sciences, Tata Institute of Fundamental Research, Bangalore 560089, India}

\author{Shasvath J. Kapadia}
\affiliation{International Centre for Theoretical Sciences, Tata Institute of Fundamental Research, Bangalore 560089, India}

\author{Parameswaran Ajith}
\affiliation{International Centre for Theoretical Sciences, Tata Institute of Fundamental Research, Bangalore 560089, India}
\affiliation{Canadian Institute for Advanced Research, CIFAR Azrieli Global Scholar, MaRS Centre, West Tower, 661 University Ave., Suite 505, Toronto, ON M5G 1M1, Canada}

\date{\today}

\begin{abstract}
We propose a method to constrain the variation of the gravitational constant $G$ with cosmic time using gravitational-wave (GW) observations of merging binary neutron stars. The method essentially relies on the fact that the maximum and minimum allowed masses of neutron stars at a particular cosmic epoch has a simple dependence on the value of $G$ at that epoch. GWs carry an imprint of the value of $G$ at the time of the merger. Thus, if the value of $G$ at merger is significantly different from its current value, the masses of the neutron stars inferred from the GW observations will be inconsistent with the theoretically allowed range. This enables us to place bounds on the variation of $G$ between the merger epoch and the present epoch. Using the observation of the binary neutron star system GW170817, we constrain the fractional difference in $G$ between the merger and the current epoch to be in the range $-1 \lesssim \Delta G/G \lesssim 8$. Assuming a monotonic variation in $G$, this corresponds to a bound on the average rate of change of $-7 \times 10^{-9}~\mathrm{yr}^{-1} \le \dot{G}/G \le 5 \times 10^{-8}~\mathrm{yr}^{-1}$ between these epochs. Future observations will put tight constraints on the deviation of $G$ over vast cosmological epochs not probed by other observations.
\end{abstract}
\maketitle

\paragraph{Introduction:---}\label{sec:introduction}

P.~A.~M.~Dirac was the first to conjecture the possibility of variation of fundamental ``constants'' of Nature with cosmic time \cite{Dirac2, Dirac1} (see also \cite{Jordan:1949zz,Jordan:1959eg,1937NW.....25..513J}). Since then, a number of alternative theories of gravity, e.g., scalar-tensor theories like Brans-Dicke theory, that predict a time varying gravitational ``constant'' $G$ \cite{BransDicke, Uzan:2010pm}, have been constructed. In general, most theories of gravity that violate the strong equivalence principle predict that the value of the gravitational constant varies with cosmic time~\cite{Will:2014kxa}. 

There are several observational bounds on the time variation of $G$, constraining its value at various cosmological epochs~\cite{Uzan:2010pm}. These include constraints derived from comparing the observed abundance of light elements with the abundance predicted by big bang nucleosynthesis models \cite{Copi:BBN, Bambi:BBN,Alvey:2019ctk}, those derived from the shape of the angular power spectrum of the cosmic microwave background \cite{Wu:2009zb}, from the light curves of type Ia supernovae \cite{Gatza:SN}, from the non-radial pulsations (asteroseismology) of white dwarfs \cite{Benvenuto:WD}, from the timing~\cite{Kaspi:1994hp} and the observed mass distributions~\cite{Thorsett} of binary pulsars, from the age of globular clusters~\cite{DeglInnocenti:1995hbi}, from helioseismology \cite{Guenther:helio}, and from monitoring the orbits of solar system planets \cite{sstest1, sstest2, sstest3} and the Earth-Moon system through radar ranging~\cite{Anderson:1992} and lunar laser ranging (LLR)~\cite{Williams:2004qba, Hofmann_2018}. In this \emph{Letter} we show how the observation of gravitational-wave (GW) signals from binary neutron stars can be used to constrain the time evolution of $G$. Assuming a monotonic variation in $G$, we also produce observational constraints on the time variation of $G$ from the binary neutron star observations by LIGO and Virgo~\cite{GW170817-DETECTION, GW190425-DETECTION}. To the best of our knowledge, this is the first such constraint placed that uses the GW signal from a binary neutron star system. It is also fourteen orders of magnitude better than the existing GW based constraint (from constraints on the dephasing of binary black hole waveforms)~\cite{YagiYunesPretorius}. We also show that the next generation of GW detectors will improve these constraints by several orders of magnitude. 

\begin{figure}
		\includegraphics[width=1\linewidth]{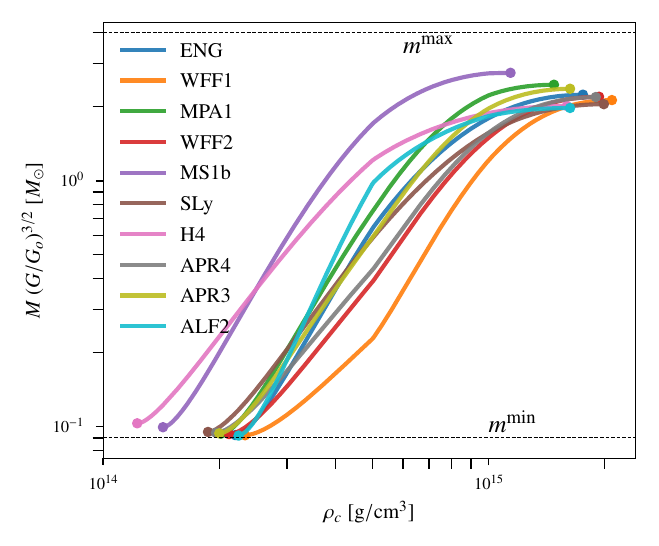}
		\caption{The mass of non-spinning neutron stars as a function of the central density $\rho_c$. These are computed by numerically solving the equations of hydrostatic equilibrium in general relativity assuming different equations of state (shown in the legend). We rescale the mass by $(G/\Go)^{3/2}$ where $G$ is the value of the gravitational constant used to solve the equations and $\Go = 6.67 \times 10^{-11}~\mathrm{m}^3\,\mathrm{kg}^{-1}\,\mathrm{s}^{-2}$ the current value of $G$. These relations between the rescaled mass and central density are independent of the value of $G$ used to compute the mass. The dots at the end of the curves correspond to the maximum/minimum mass supported by each equation of state. The horizontal dashed lines correspond to the maximum/minimum possible mass of neutron stars that we assume in this paper, independent of the equation of state.}
		\label{fig:M_R_plot}
\end{figure}

\paragraph{Neutron star mass limits and the gravitational constant:---} The mass of a spherically symmetric star is determined by the following equations of hydrostatic equilibrium 
\begin{eqnarray} 
\frac{dm(r)}{dr} = 4 \pi r^2 \rho(r) \, , ~~~ \frac{dP(r)}{dr}  =  \frac{-G \, m(r) \, \rho(r)}{r^2} \, \mathcal{C}(r),  
\label{eq:hydro_eq}
\end{eqnarray} 
where $m(r)$ denotes the (gravitational) mass enclosed by a radius $r$ while $P(r)$ and $\rho(r)$ denote the pressure and density at $r$. The dimensionless quantity $\mathcal{C}(r)$ denotes the relativistic corrections. That is, $\mathcal{C}(r) = 1$ in Newtonian gravity, $\mathcal{C}(r) = \left[ 1 + P(r)/\rho(r) c^2 \right] \, \left[1 + 4 \pi r^3 P(r)/m(r) c^2 \right] \left[1 - 2 G m(r)/c^2 r\right]^{-1}$ in general relativity~\cite{tov}, while $\mathcal{C}(r)$ depends on additional quantities such as scalar fields in alternative theories of gravity (see, e.g.,~\cite{Damour:1993hw}). The mass of the star can be computed by solving Eq.~\eqref{eq:hydro_eq} along with an equation of state that relates the density and pressure. 

A dimensional analysis of Eq.~\eqref{eq:hydro_eq} reveals that the mass of the star scales as $G^{-3/2}$. This is illustrated in Fig.~\ref{fig:M_R_plot}, which shows the mass of non-spinning neutron stars as a function of the central density, computed by numerically solving the equations of hydrostatic equilibrium in general relativity~\cite{tov}.  We rescale the mass by $(G/\Go)^{3/2}$, where $G$ is the value of the gravitational constant used to solve the equations and $\Go = 6.67 \times 10^{-11}~\mathrm{m}^3\,\mathrm{kg}^{-1}\,\mathrm{s}^{-2}$ the current value of $G$. The rescaled mass as a function of the central density is independent of the value of $G$ used to compute the mass. In particular, for a given equation of state, the central densities corresponding to the maximum and minimum masses are independent of $G$. 

We entertain the possibility that the value of the gravitational constant during the merger, $G_s$, could be different from its current value $\Go$~\cite{Wu:2019pbm}. Following common practice~\cite{Uzan:2010pm,Will:2014kxa}, we assume that all other fundamental constants and fundamental interactions (and hence the nuclear equation of state) remain unchanged across cosmic time~\footnote{A variation in $G$ should also affect the nuclear reaction rates and hence the stellar evolution. This will alter the rate of compact binary mergers at various cosmological epochs. We neglect this effect while estimating the expected constraints using upcoming detectors.}. It is well known that neutron stars have a minimum (maximum) mass limit~\cite{Lattimer:2004pg} below (above) which the neutron star will gravitationally unbind (collapse). Fig.~\ref{fig:M_R_plot} shows that these mass limits scale as $G^{-3/2}$. Thus, the mass limits on the neutron star at the time of the merger are given by 
\begin{equation}
\ms^\mathrm{min,~max} = m^\mathrm{min,~max} ~ (\Gs/\Go)^{-3/2}
\label{eq:m_max_min}
\end{equation}
where $m^\mathrm{min,~max}$ are the minimum and maximum mass limit computed using the current value of the gravitational constant, $\Go$. The precise values of $m^\mathrm{min,~max}$ will depend on the nuclear equation of state, which, to date, remains an open question. Nevertheless, guided by the plethora of available equation of state models, we can make a conservative choice of $m^\mathrm{min} = 0.09~M_\odot$~\cite{Haensel:2002cia} and $m^\mathrm{max} = 4~M_\odot$ (shown by horizontal dashed lines in Fig.~\ref{fig:M_R_plot}), where the maximum mass includes the effects of maximal rotation~\cite{Koranda:1996jm,Rhoades:1974fn}. 

\paragraph{Constraining the cosmic evolution of the gravitational constant:---}  

The GW signal produced by the inspiral of a binary neutron star system can be written in the frequency domain as $h(f) = A(f) \exp {i \, \Psi(f)}$, where $A(f)$ is the amplitude, $\Psi(f)$ is the phase and $f$ is the Fourier frequency. The post-Newtonian expansion of the phase is given by~\cite{BIOPS}  
\begin{equation}
\Psi(f) = \frac{3}{128 \, \eta \, v^5} ~ \sum_k \psi_k(\blambda) \, v^k. 
\label{eq:gwphase}
\end{equation}
Here, $v := \left({\pi \, \Gs \Ms f}/{c^3}\right)^{1/3}$ is a dimensionless parameter,  where $\Ms = m_\mathrm{s,1} +  m_\mathrm{s,2}$ is the total mass of the binary, while $\psi_k$ are some dimensionless coefficients that depend on a set of intrinsic parameters $\blambda$ of the binary, such as the symmetric mass ratio $\eta$ and dimensionless spins of the neutron stars, but \emph{not} on the value of $G$~\footnote{The parameters that appear in the waveform describing the effects of the internal structure of the neutron stars (e.g., spin- and tidal-induced deformations) \emph{do} depend on $G$. However, their (dimensionless) contributions to the waveform are $G$-independent, similarly to how they are not affected by the cosmological redshift~\cite{Messenger:2011gi}. 
Additionally, the phase of the GW also depends on extrinsic parameters such as the location and orientation of the binary, which do not affect the current discussion.}. 

The detection of GWs from compact binaries, and the estimation of source parameters is performed by phase matching the data with theoretical templates of the expected signal. Typically, several hundreds of thousands of templates, corresponding to different values of source parameters $\blambda$, need to be correlated with the data. These templates are generated using the current value $\Go$ of the gravitational constant. That is, templates use $v := \left({\pi \, \Go \Mo f}/{c^3}\right)^{1/3}$ in Eq.~\eqref{eq:gwphase}, where $\Mo$ is the total mass of the binary parametrizing a given template.

\begin{figure}[t]
		\includegraphics[width=1\linewidth]{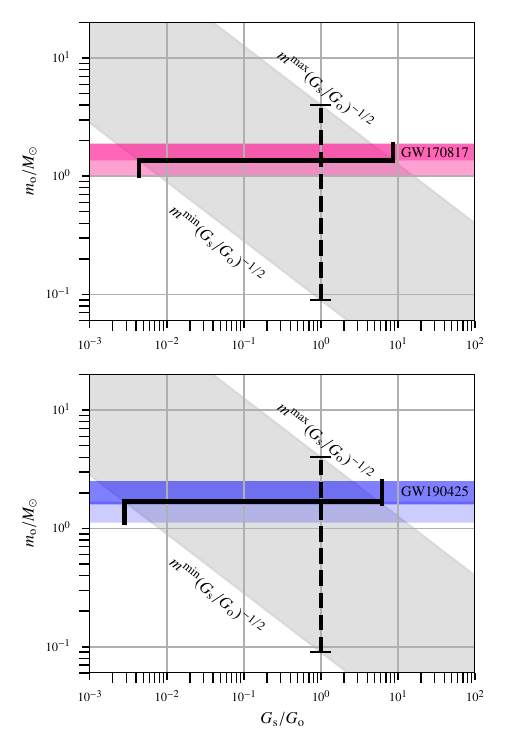}
		\caption{The horizontal axis corresponds to the assumed value of $G$ at the epoch of the merger of the neutron stars, denoted by $\Gs$ (in units of the current value $\Go$), and the vertical axis corresponds to the measured value of the neutron star mass $\mo$ (in units of solar mass) from GW observations. The horizontal patches correspond to the measured value of the neutron star masses from GW170817 and GW190425 (90\% credible regions of the marginalized posteriors on the component masses~\cite{GW170817-PE, GW190425-DETECTION}; darker shades correspond to primary masses while lighter shades correspond to secondary masses). The vertical dashed lines correspond to the theoretically allowed range of neutron star masses $m^\mathrm{min} - m^\mathrm{max}$ calculated using the current value, $\Go$, of the gravitational constant. The tilted gray regions show the allowed range of the observed neutron star mass; i.e., $m^\mathrm{min}~(\Gs/\Go)^{-1/2}$ to $m^\mathrm{max}~(\Gs/\Go)^{-1/2}$, where $m^\mathrm{min} = 0.09 M_\odot$ and $m^\mathrm{max} = 4 M_\odot$. The horizontal error bars show the constraints on $\Gs/\Go$ that we obtain by requiring that the measured values overlap with the theoretically allowed (gray) region.}
		\label{fig:Gf_by_Go}
\end{figure}

\begin{figure*}[t]
	\centering
	\includegraphics[width=0.95\linewidth]{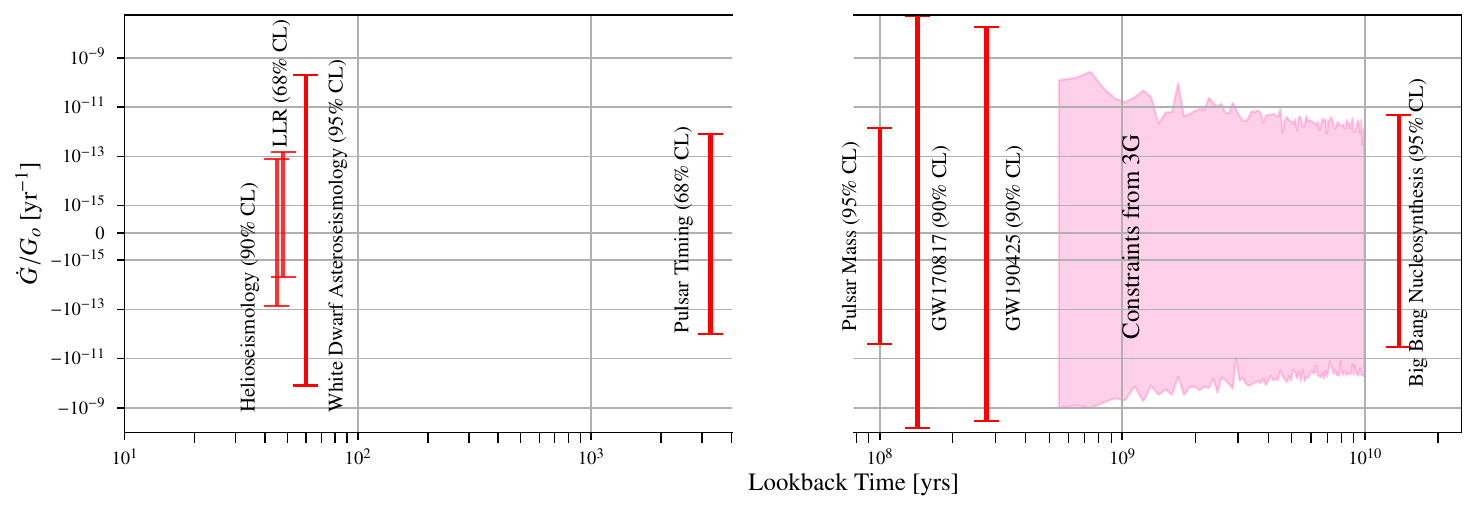}
	\caption{Comparison of the constraints on $\dot{G}/\Go$ obtained from various observations with confidence levels of the constraints indicated in brackets (LLR denotes constraints from lunar laser ranging). The horizontal axis shows the look back time. The constraints presented this paper are labeled as GW170817 and GW190425, while the expected constraints from 10 year of observations with third-generation (3G) GW detectors is shown by the pink band.}
	\label{fig:gdot_plots}
\end{figure*}

Template matching is achieved when the phase of the template is matched with the phase of the signal, which, in turn, occurs when the set of parameters $\blambda$ of the templates matches that of the source. It can be seen from Eq.~\eqref{eq:gwphase} that, in addition, the following condition also has to be satisfied (neglecting cosmological redshift)
\begin{equation}
\left(\frac{\pi\, \Go \Mo f}{c^3}\right)^{1/3} = \left(\frac{\pi\, \Gs \Ms f}{c^3}\right)^{1/3},
\label{eq:templmatching}
\end{equation}
Thus, when $\Gs \neq \Go$, the estimated total mass of the binary from this template matching (and hence the masses $m_\mathrm{o,1}$ and $m_\mathrm{o,2}$ of the neutron stars) will be biased. 
\begin{equation}
\Mo = \frac{\Gs}{\Go} \Ms ~~~~ \Longrightarrow ~~~~ \mo = \frac{\Gs}{\Go} \ms, 
\label{eq:m_obs}
\end{equation}
where $\ms$ and $\mo$ are the actual and estimated masses of the neutron stars, respectively. We have omitted the subscripts 1, 2 for simplicity. Due to the cosmological expansion, GWs will be redshifted. The effect of this can be captured by applying the redshift factor $(1+z)$ to the masses, i.e., by replacing $m$ by $m(1+z)$~\cite{Cutler:1994ys}. Due to this effect, Eq.\eqref{eq:m_obs} will be modified to  
\begin{equation}
\mo = \frac{\Gs}{\Go} \, \left(\frac{1+\zs}{1+\zo}\right) \, \ms, 
\label{eq:m_obs_with_z}
\end{equation}
where $\zs$ is the true cosmological redshift of the source and $\zo$ is the estimated redshift --- either from a direct electromagnetic observation of the host galaxy or by converting the luminosity distance estimated from the GW observation, assuming a cosmology. Using Eq.~\eqref{eq:m_max_min} in Eq.~\eqref{eq:m_obs_with_z} gives the maximum and minimum values of observable masses 
\begin{equation}
\mo^\mathrm{max,~min} = m^\mathrm{max,~min}~(\Gs/\Go)^{-1/2}~\left(\frac{1+\zs}{1+\zo}\right). 
\label{eq:m_obs_fn_Gf_z}
\end{equation}

If the redshift is accurately known from an electromagnetic observation (as in the case of the neutron star merger event GW170817~\cite{GBM:2017lvd}), then $\zo = \zs$, and Eq.\eqref{eq:m_obs_fn_Gf_z} reduces to 
\begin{equation}
\mo^\mathrm{max,~min} = m^\mathrm{max,~min}~(\Gs/\Go)^{-1/2}. 
\label{eq:m_obs_fn_Gf}
\end{equation}
That is, when $\Gs \neq \Go$, the allowed range of neutron star masses (as measured from GW observations) will be different from the range $m^\mathrm{min}$ --- $m^\mathrm{max}$. This is shown by the gray region in Fig.~\ref{fig:Gf_by_Go}. If $\Gs$ has a significant deviation from the current value $\Go$, the observed values of neutron star masses (from GW observations, shown as horizontal bands in Fig.~\ref{fig:Gf_by_Go}) would have gone out of the allowed range of neutron star masses (gray region). Hence such large deviations are ruled out by current GW observations. 

Even if the redshift is not known from electromagnetic observations, the luminosity distance estimated from GW observations can be used to infer the redshift assuming a cosmological model. However, if $G$ varies over cosmological time, then the inferred redshift will have a bias because the cosmological model uses the current value, $\Go$, of $G$. The allowed mass range of neutron stars should be made larger to accommodate this effect (see the parenthetical ratio of redshifts in Eq.\eqref{eq:m_obs_fn_Gf_z}). This effect is negligible for GW190425 thanks to its relatively low distance; however we will need to consider this effect for binary neutron star observations from large distances (for e.g., by the next generation detectors).

\paragraph{Results:---}
Figure~\ref{fig:Gf_by_Go} summarizes the constraints on the variation of $G$ between the epoch of the merger of the neutron stars and the present epoch, obtained from the two putative binary neutron star observations by LIGO and Virgo, GW170817~\cite{GW170817-DETECTION} and GW190425~\cite{GW190425-DETECTION}. The horizontal axis corresponds to the assumed value of $G$ at the epoch of the merger, denoted by $\Gs$ (in units of the current value $\Go$), and the vertical axis corresponds to the neutron star mass $\mo$ (in units of solar mass) that we would estimate from GW observations. The tilted gray regions show the allowed range of the observed neutron star mass; i.e., $m^\mathrm{min}~(\Gs/\Go)^{-1/2}$ to $m^\mathrm{max}~(\Gs/\Go)^{-1/2}$, where $m^\mathrm{min} = 0.09 M_\odot$ and $m^\mathrm{max} = 4 M_\odot$. The two horizontal bands correspond to the measured value of the neutron star masses (90\% credible regions of the marginalized posteriors on the component masses from LIGO-Virgo observations~\cite{GW170817-DETECTION,GW190425-DETECTION}).  

When {$\Gs \lesssim 4\times 10^{-3} \Go$ or when $\Gs \gtrsim 9 \Go$}, the observed mass range of the neutron stars in GW170817 will go out of the predicted mass range. Thus, the fractional deviation $\Delta G/\Go := (\Gs - \Go)/\Go$ in gravitational constant is constrained to be {$-0.996 \lesssim \Delta G/\Go \lesssim 8$}. Similarly, GW190425 constrains $\Gs$ to be in the range {$-0.997 \lesssim \Delta G/\Go \lesssim 5$}. Note, however, that $\Gs$ in different observations will refer to different epochs (that of the individual mergers). 

If we know the time $\Delta t$ elapsed between the merger and the GW observation, we can compute an average rate of change of the $G$ during this period: $\dot{G}/\Go \simeq \Delta G/(\Go \Delta t)$. Since the luminosity distance to the source is measured, by assuming a cosmology, we can compute the time taken by the signal to reach the observer from the source~\footnote{Note that a time varying $G$ will also change the cosmology, and hence the GW propagation time. For the events we analyze, this does not change the results by more than a factor of a few, and hence we neglect it. We assume standard values for cosmological parameters as quoted in \cite{planck15}.}. GW170817 and GW190425 provide the constraints {$-7 \times 10^{-9}~\mathrm{yr}^{-1} \lesssim \dot{G}/\Go \lesssim 5 \times 10^{-8} ~\mathrm{yr}^{-1} $} and  {$-4 \times 10^{-9} ~  \mathrm{yr}^{-1} \lesssim \dot{G}/\Go \lesssim 2 \times 10^{-8} ~\mathrm{yr}^{-1} $}, respectively. These are plotted in Fig.~\ref{fig:gdot_plots}, along with similar constraints obtained from other astronomical observations. 

The proposed third generation detectors will detect millions of binary neutron star mergers out to cosmological distances~\cite{Mills:2017urp}. These will include binaries with neutron star masses close to their theoretical upper/lower limits, providing tight constraints on the value of $G$ at the corresponding cosmological epochs. In order to characterize the expected constraints from such observations, we simulate populations of neutron star mergers assuming three different mass distributions. The first is uniform in component masses in the range $1-3 M_\odot$; the second  \cite{Farrow:2019xnc} and third \cite{Kiziltan:2013oja} are based on observed Galactic binary pulsars. We assume a 10 year observing run of a network of two Cosmic Explorer~\cite{CE} detectors at the LIGO Hanford and Livingston sites, and an Einstein Telescope~\cite{Sathyaprakash:2012jk}  at the Virgo site. Using the redshift distribution from \cite{Mills:2017urp} and assuming a local merger rate of $320$ Gpc$^{-3}$ yr$^{-1}$ \cite{Abbott:2020gyp}, this will amount to $\sim 2.5$ million detections,  out to redshift of 1.7. 

\begin{figure}
	\centering
	\includegraphics[width=1.0\linewidth]{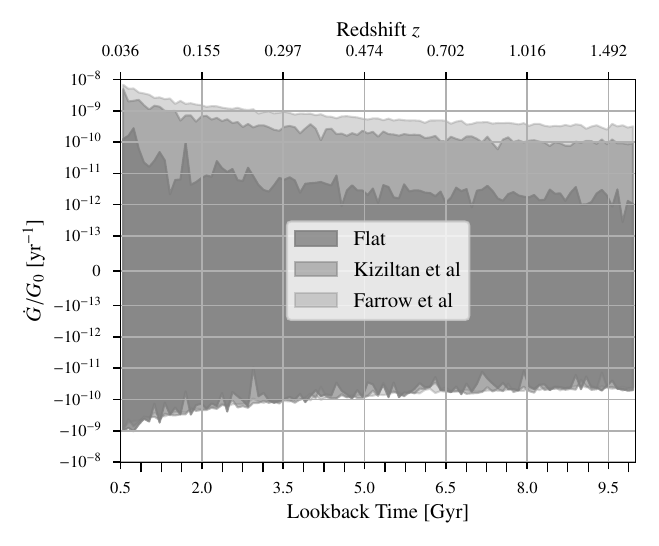}
	\caption{Expected constraints on $\dot{G}/\Go$ from 10 year observations of third-generation GW detectors. We assume three different mass distributions of neutron stars, and that $\sim 1\%$ of the mergers will have a detectable electromagnetic counterpart from which the cosmological redshift can be estimated.} 
	\label{fig:gdot_3G}
\end{figure}

We assume that  $\sim 1\%$ of these mergers will produce detectable short gamma-ray bursts \cite{MetzgerBergerSGRB}, which will enable a direct measurement of source redshifts $\zs$ using follow-up observations.  This allows us to use Eq.~\eqref{eq:m_obs_fn_Gf} instead of Eq.~\eqref{eq:m_obs_fn_Gf_z}. We estimate the expected errors on measuring the component masses of these 25,000 binaries using the \textsc{GWBench} \cite{Borhanian:2020ypi} code. The best constraints on $\dot{G}/\Go$ from different cosmological epochs (bins of $0.096$ Gyr) are plotted in Fig.~\ref{fig:gdot_3G} as a function of lookback time. For this, we assume a more optimistic choice of $1-3 M_\odot$ for the theoretical mass limits of neutron stars; a conservative choice of $0.09-4 M_\odot$ will worsen these constraints  by a factor of $5-10$. As expected, we get the best constraints when the observed mass range span the entire theoretically allowed mass range --- for e.g., in the case of the flat mass distribution considered here (also quoted in Fig.~\ref{fig:gdot_plots}). Over a wide cosmological epoch spanning about 10 Gyrs, GW observations will, and perhaps are the only way to, probe the variation of $G$.~\footnote{Even binary neutron star mergers without electromagnetic counterparts can be used to put constraints, assuming specific cosmological models with varying $G$ (e.g,~\cite{Barrow:2005hw}). Note also that multi-band observations (using ground- and space- based detectors) of binary black holes are also expected to produce comparable constraints~\cite{Perkins2021}.}

\paragraph{Summary and outlook:---} 
We proposed a method to constrain the evolution of the gravitational constant over cosmic time using GW observations of binary neutron stars, and produced the first constraints with LIGO-Virgo observations (Fig.~\ref{fig:Gf_by_Go}). Although these constraints are not as tight as the best available bounds from other measurements, they sample a different cosmological epoch that is not covered by other observations. Additional detections of binary neutron stars would constrain the variation of $G$ at different epochs.  

Gravitational-wave observations sampling the extremes of the neutron star masses will further tighten these bounds. For example, the measurement of a high-mass (low-mass) neutron star will constrain the possibility of $\Gs > \Go$ ($\Gs < \Go$). If there are multiple observations from similar distances (i.e., mergers from similar cosmological epochs), these constraints can be combined, thus further tightening them. Tighter theoretical mass limits for neutron stars (through better understanding of the nuclear equation of state) will also help us to place tighter bounds on $\Delta G$. By observing a large number of binary neutron stars (some sampling the extremes of the neutron star mass limit), the upcoming generation of GW detectors will be able to tightly constrain the evolution of $G$ over vast cosmological epochs spanning $\sim10$ Gyrs (Fig.~\ref{fig:gdot_plots}).

Note that this test relies on coalescing binaries having at least one neutron star, and is therefore extendable to neutron star - black hole mergers as well, especially if these can be associated with an electromagnetic counterpart. We have assumed that the GW events GW170817 and GW190425 are binary neutron star mergers. Although the possibility of GW190425 being a binary black hole can not be ruled out~\cite{GW190425-DETECTION}, the observed electromagnetic counterparts of GW170817 strongly suggests that the binary contains at least one neutron star. 
In principle, comparison of the predicted luminosity of the electromagnetic counterparts such as kilonovae with observations will also allow us to constrain the time evolution of $G$. This requires accurate models of the electromagnetic emission from binary mergers, which might become available in the near future. 


\paragraph{Acknowledgments:---} 
{We are grateful to Emanuele Berti and the anonymous referee for the careful review and valuable feedback.} We also thank Nathan K. Johnson-McDaniel, Ajit Kumar Mehta and Bala Iyer for very useful comments on the manuscript. We thank K. G. Arun, Arif Shaikh, Apratim Ganguly, Kanhaiya Lal Pandey, Srashti Goyal, Soummyadip Basak, Mukesh Kumar Singh, Souvik Jana, Uddeepta Deka, Sanskriti Chitransh, B. S. Sathyaprakash and Thomas Dent for illuminating discussions. Our research is supported by the Department of Atomic Energy, Government of India. In addition, SJK's research was supported by the Simons Foundation through a Targeted Grant to the International Centre for Theoretical Sciences (ICTS). PA's research was supported by the Max Planck Society through a Max Planck Partner Group at ICTS-TIFR and by the Canadian Institute for Advanced Research through the CIFAR Azrieli Global Scholars program. 

\bibliographystyle{apsrev-nourl}
\bibliography{references}

\end{document}